\documentclass[a4paper,fleqn,usenatbib]{mnras}

\usepackage[T1]{fontenc}
\usepackage{ae,aecompl}

\usepackage{graphicx}	% Including figure files
\usepackage{amsmath}	% Advanced maths commands
\usepackage{amssymb}	% Extra maths symbols

\title[Milky Way bulge globular cluster candidates]{On the physical nature of globular cluster candidates in the Milky Way bulge}

\author[Andr\'es E. Piatti]{
A.E. Piatti$^{1,2}$\thanks{E-mail: andres@oac.unc.edu.ar}
\\
$^{1}$Consejo Nacional de Investigaciones Cient\'{\i}ficas y T\'ecnicas, Av. Rivadavia 1917, 
C1033AAJ, Buenos Aires, Argentina\\
$^{2}$Observatorio Astron\'omico, Universidad Nacional de C\'ordoba, Laprida 854, 5000, 
C\'ordoba, Argentina\\
}

\date{Accepted XXX. Received YYY; in original form ZZZ}

\pubyear{2017}

\begin{document}
\label{firstpage}
\pagerange{\pageref{firstpage}--\pageref{lastpage}}
\maketitle

\begin{abstract}
We present results from 2MASS $JK_s$ photometry on the physical reality of
recently reported globular cluster (GC) candidates in the Milky Way (MW) bulge.
We relied our analysis on photometric membership probabilities
that allowed us to distinguish real stellar aggregates from the composite
field star population. When building colour-magnitude diagrams and
stellar density maps for stars at different membership probability levels, the
genuine GC candidate populations are clearly highlighted. We then used
the tip of the red giant branch (RGB) as distance estimator, resulting
heliocentric distances that place many of the objects in regions near 
of the MW bulge
where no GC had been previously recognised. Some few GC candidates
resulted to be MW halo/disc objects. Metallicities estimated from the
standard RGB method are in agreement with the values expected according to the
position of the GC candidates in the Galaxy.
Finally, we derived from the first time their structural parameters.
We found that the studied objects have core, half-light and
tidal radii in the ranges spanned by the population of known MW GCs.  Their
internal dynamical evolutionary stages will be described properly when their
masses are estimated.

\end{abstract}

\begin{keywords}
techniques: photometric -- galaxies: individual: Milky Way --
galaxies: star clusters: general 
\end{keywords}

%%%%%%%%%%%%%%%%%%%%%%%%%%%%%%%%%%%%%%%%%%%%%%%%%%

%%%%%%%%%%%%%%%%% BODY OF PAPER %%%%%%%%%%%%%%%%%%

\section{Introduction}

Recently, \citet[][hereafter M17]{minnitietal2017} have reported 22 new globular 
cluster (GC) candidates from overdensities in red giant branch (RGB) star maps 
built from photometric data sets of the  VISTA\footnote{Visible and Infrared 
Survey Telescope for Astronomy} Variables in the V\'{\i}a L\'actea (VVV) survey
 \citep{minnitietal2010,saitoetal2012}. 
These candidates are projected in direction towards the Milky Way (MW) bulge  (10$^o$ $\le$ 
{\it l} $\le$ 350$^o$) and were assumed to be as old as the ancient MW 
GCs. However, main sequence turnoffs of theoretical isochrones 
for an age of log($t$ yr$^{-1}$) = 10.1 (12.6 Gyr) overplotted on 
the GC colour-magnitude diagrams (CMDs) constructed by M17 (see their figure 3) 
are $\sim$ 0.5-1.5 mag fainter than the observed ones. We checked these
values by first 
snapping 12 M17's CMDs with clear GC sequences ($\#$3, 5, 6, 8, 15, 16, 17, 18, 19, 20, 21, 22), then overplotted the respective \citet{betal12}'s 
isochrones using {\sc python} tools and finally estimated them by visual inspection.
The isochrones for the metallicties of the 
corresponding reference 
GCs assigned by M17 were previously shifted by
reddening and distance using the values obtained by them (their table 1).
Indeed, the main sequence turnoffs of the well-known
GCs NGC\,6637 and 6642, also included in their CMDs sample, are fainter than the limiting 
magnitude reached by the VVV survey. At this point, it results also a little bit curious
that most of the star field decontaminated CMDs present well-isolated  main sequences 
turnoffs in comparison with the upper part of these CMDs, where residual stars outside the 
RGBs are seen. 

Such mismatches trigger the unavoidable question about the reliability of the distance 
estimates for these GC candidates, the assumption of ages similar to those of old MW bulge
GCs, and even about the real nature of such RGB star overdensities, among others. M17 estimated 
GC candidate distances by comparing the position of the red clump in their CMDs with those 
of six old known MW GCs also observed by the VVV survey. 
GC candidates with clear red clumps were considered metal-rich  ([Fe/H] $>$ -1 dex) objects, while
those without prominent ones were treated as metal-poor ([Fe/H] $<$ -1 dex) candidates. 
Such a distinction was necessary to select the reference MW GCs to which measure the difference in
red clump magnitudes. M17 mentioned that, if they were real GCs, they should 
be grouped more or less symmetrically around the Galactic centre.

We here embarked in a thorough study aiming at confirming the physical nature of these objects
from independent data sets and analysis approaches.
We relied on the extensive use of the tip of the red giant branch (RGBT) 
to determine
extragalactic distances 
\citep[see, e.g.,][and references therein]{serenellietal2017,madoreetal2018,hoytetal2018}, to
estimate the GC candidate distances
making use the  2MASS database \citep{setal2006}.
While the VVV
photometry saturates for magnitudes brighter than $K_s$ $\sim$ 12.0 mag, the 2MASS one is able to reach
well above the RGBT of MW bulge GCs 
\citep[$K_s$ $\sim$ 8.0 mag,][]{cohenetal2015}. Therefore, from 2MASS photometry 
is possible to trace the upper part of the RGBs of the M17's GC candidates, and by employing the
magnitude of the RGBT  to estimate reliable distances  \citep{mulleretal2018,sabbietal2018,beatonetal2018}. 
It is worth mentioning that 2MASS
data have been extensively used in the last decades to search for new GCs \citep[see, e.g.,][]{hurtetal2000,ivanovetal2000,ivanovetal2002,borissovaetal2003,froebrichetal2005,bicaetal2007,bonattoetal2007,straderetal2008,scholzetal2015}
.

In Section 2 we describe the procedure applied to assign
membership probabilities to stars observed in the GC candidate fields, the technique to estimate
their metal content, and the methods for constructing their stellar radial profiles and 
estimating their structural parameters. In Section 3, we discuss the results to the light of our
previous knowledge of the MW bulge GC population and those of M17. 
 Particularly, we compared
the derived structural parameters (core, half-light, tidal radii) with those compiled by 
\citet{harris1996} and our distance estimates with those derived by M17.
Finally, in Section 4 we summarize the main conclusions of this work.

\section{Data handling}

We downloaded 2MASS $JK_s$ data sets from the IPAC Infrared Science 
Archive\footnote{www.ipac.caltech.edu/2mass/overview/access.html} (IRSA)
for circular regions with radii of 30 arcmin around the GC candidate centres. 
Such regions pretty well cover the whole objects extensions and their surrounding
fields. Indeed, the six
MW GCs used as reference by M17 have radii between $\sim$ 4 - 7 arcmin \citep{harris1996}, 
while M17 estimated sizes of $\sim$ 2 - 5 arcmin in radius for their 22 GC candidates.  

We first defined a circular region with a radius of 10 arcmin around each GC candidate centre
and eight additional circular regions of that same size,
uniformly distributed around its centre, at a distance of 20 arcmin. For the eight
GC candidate surrounding regions, we built their respective $K_s$ versus $J-K_s$ CMDs and 
produced a collection of rectangular cells centred at the position of each star and with 
dimensions  - defined independently for magnitude and colour - in such a way that there is 
no star inside each cell but that at its centre, with the following closest one located
at one of the four corners of the rectangular cell. This array of cells properly
reproduces the stellar density of that surrounding field and appropriately traces 
its luminosity function and colour distribution in the CMD. We refer to the works by 
\citet{petal2018}, among others, where the reader will
find illustrations about the generation of the rectangular cells. The method was developed by 
 \citet{pb12} with the aim of providing a robust approach to 
statistically clean the CMDs from the field star contamination. We have used it since then
dealing with optical and near-infrared photometries of MW and Magellanic Clouds star clusters in
crowded and highly reddened stars fields \citep[see, e.g.][]{pc2017}. 

\begin{figure*}
\includegraphics[width=\textwidth]{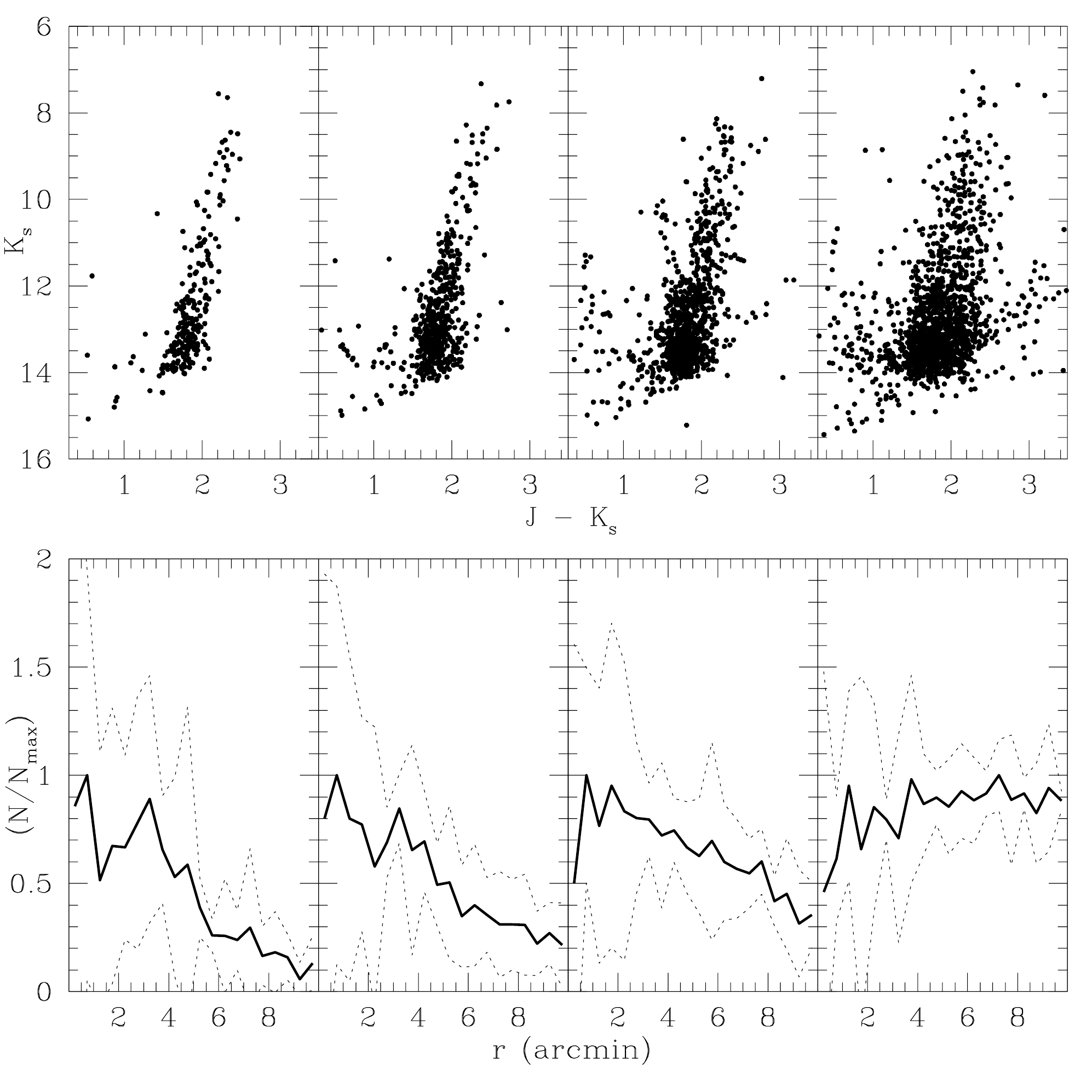}
    \caption{{\it Top:} $K_s$ versus $J-Ks$ CMDs for stars distributed in the field of Minni\,16
for membership probabilities $P$ = 100, 88, 75 and 50\% (from left to right), respectively.
{\it Bottom:} Observed stellar radial profiles for the above respective $P$ values. The solid
and dotted lines represent the average and standard deviation. Similar figures for the remaining 
21 GC candidates are provided as supplementary material in the online version of the journal.
}
   \label{fig:fig1}
\end{figure*}

Each collection of cells per surrounding field CMD was superimposed on the
GC candidate CMD, and then subtracted from it one star per defined cell: the closest
one to the cells' centres. Thus we cleaned the GC candidates CMDs by eliminating the
same number of stars as in the surrounding field CMD, according to the particular
pattern along the magnitude and colour ranges of that surrounding field. Since we repeated 
this exercise eight times - one per surrounding field CMD -, we generated eight different
cleaned GC candidate CMDs. We then combined all of them by producing a master
GC candidate cleaned CMD where we assigned to each star a statistical photometric membership status.
Stars that were kept unsubtracted eight times have a membership probability $P$ = 100\%;
those stars left in the master cleaned CMD seven times, a $P$ value of 88\%; stars six times unsubtracted 
have $P$= 75\%, and so forth, until $P$ = 13\% for those stars that appear once in the master cleaned CMD. 
Thus, we were able to recognise stellar populations that
most probably belong to the field - those having the smallest $P$ values -, i.e., stars 
that are more or less similarly distributed as a function of stellar density, luminosity 
function and colour distribution throughout the GC candidate field; and stars that represent 
a local intrinsic feature of that part of the sky - stars with the largest $P$ values -. 
These stars could be part of a star cluster or of an apparent enhancement of stars
along that line-of-sight. Stars with $P$ = 50\% can be either part of the composite field
population or belong to the local stellar enhancement.
Fig.~\ref{fig:fig1} depicts some resulting CMDs of Minni\,16 for different membership 
probabilities. CMDs for smaller $P$ values ($P$ = 38, 25 and 13\%) are less worth for our 
study, since they show repeatedly the composite field population.
 As can be seen,
the broadnesses of the RGBs become narrower as the CMDs are built with larger $P$ values, suggesting
the transition from a composite field stellar population to that of a stellar aggregate.

\citet{serenellietal2017} have comprehensively studied the characteristics of the  RGBT
from a theoretical point of view and provided a set of relationships to compute
the absolute $K_s$ magnitude of the  RGBT ($M_{K_s}^{RGBT}$) in terms of the intrinsic 
$(J-K_s)_o$ colour (not affected by reddening). We took advantage of these relationships
to estimate the GC candidate distances. We first measured the observed $K_s$ magnitudes
and $J-K_s$ colours of the RGBT in the GC candidate cleaned CMDs; then we corrected them by
reddening effect using the $E(J-K_s)$ values in M17 and the relation  $A_{K_s}$ = 
0.428$E(J-K_s)$ \citep{alonsogarciaetal2015}. Finally, we computed the 
 $M_{K_s}^{RGBT}$
absolute magnitudes
from the dereddened $(J-K_s)_o$ colours and the equations of 
\citet[][see their table 1]{serenellietal2017}, and the absolute distance
 moduli using 
these latter values and the
dereddened $K_s$ magnitudes. Because CMDs for $P$ = 88\% and 100\% exhibit
narrower RGBs respect to CMDs of smaller $P$ values, we used both of them to measure
$K_s$ magnitudes and $J-K_s$ colours of the RGBTs. We built the $K_s$ luminosity function 
of the RGB using bins of 0.1 mag wide and then identified by visual inspection
its sudden truncation, as well as the corresponding $J-K_s$ colour. 
In some few cases we also used
 the $P$= 75\% CMDs.  We estimated errors of  $\sigma$($K_s^{RGBT}$) = 0.15
mag and $\sigma$($(J-K_s)^{RGBT}$) = 0.05 mag, respectively, to which we
 added the 
uncertainties in $E(J-K_s)$ ($\sigma$$E(J-K_s)$ = 0.05 mag) to calculate
 errors in
their dereddened values. Table~\ref{tab:table1} lists the measured  $K_s^{RGBT}$ magnitudes
and $(J-K_s)^{RGBT}$ colours and the resulting distances. As an external check, we
repeated this procedure to estimate the distance of NGC\,6637. We used  2MASS 
$JK_s$ photometry, cleaned its CMD and measured the magnitude and the colour of its RGBT and
obtained the same value provided by \citet{harris1996} within 0.2 kpc.

\begin{table*}
\centering
\caption{Derived properties of studied GC candidates.}
\label{tab:table1}
\begin{tabular}{@{}lcccccccc}\hline
ID & $(J-K_s)^{RGBT}$ & $K_s^{RGBT}$ &  $d$  &  [Fe/H] &  $r_c$  &  $r_h$   &  $r_t$ & Class\\
   & (mag)   & (mag)  & (kpc)   & (dex)  & (pc) & (pc) & (pc) & \\
\hline
Minni\,1  &  1.10  & 10.00  &  16.2 $\pm$ 3.0 &    -1.2  &    0.9 $\pm$ 0.5   &   5.7 $\pm$ 1.4  &  47.1 $\pm$ 14.1      & disc/halo\\    
          &        &        &  8.1     &       &   0.5 $\pm$    0.2  &   2.8  $\pm$   0.7  &   23.6  $\pm$   7.1\\
Minni\,  2  &  1.15  & 10.80  &  24.2 $\pm$ 4.5 &    -1.5   &   1.4 $\pm$ 0.7   &   8.4 $\pm$ 2.1  &   70.4 $\pm$ 21.1   & disc/halo\\ 
          &        &        &6.6      &      &    0.4  $\pm$   0.2   &   2.3 $\pm$    0.6 &    19.2 $\pm$    5.8\\
Minni\,  3  &  0.90  & 12.20  &  38.1 $\pm$ 6.7 &    -2.0   &   2.2  $\pm$ 1.1   &  13.3 $\pm$ 3.3  &  110.8 $\pm$ 33.2  & disc/halo\\ 
          &        &        &7.0     &       &    0.4  $\pm$   0.2   &   2.4  $\pm$   0.6  &   20.4  $\pm$   6.1\\
Minni\,  4  &  1.40  &  8.40  &   9.8 $\pm$ 1.9 &    -0.3  &    0.6 $\pm$ 0.3   &   3.4 $\pm$ 0.9  &   28.5 $\pm$ 8.6    & bulge\\ 
          &        &        &5.3    &        &    0.3  $\pm$   0.2  &    1.9  $\pm$   0.5  &   15.4  $\pm$   4.6\\
 Minni\, 5  &  1.70  &  7.75  &   9.8 $\pm$ 2.1 &   -0.1  &    0.6 $\pm$ 0.3   &   3.4 $\pm$ 0.9  &   28.5 $\pm$ 8.6     &bulge\\ 
          &        &        & 8.5   &         &    0.5  $\pm$   0.2  &    3.0 $\pm$    0.7 &    24.7 $\pm$    7.4\\
Minni\,  6  &  1.40  &  8.10  &   8.3 $\pm$ 1.7 &   -0.3  &    0.5 $\pm$ 0.2   &   2.9 $\pm$ 0.7  &   24.1 $\pm$ 7.2     &bulge\\ 
          &        &        & 8.4   &         &    0.5  $\pm$   0.2  &    2.9 $\pm$    0.7  &   24.4 $\pm$    7.3\\
Minni\,  7  &  1.50  &  9.00  &  14.0 $\pm$ 2.8 &   -0.3   &   0.8 $\pm$ 0.4   &   4.9 $\pm$ 1.2  &   40.7 $\pm$ 12.2    & ---\\ 
          &        &        &6.8    &         &   0.4  $\pm$   0.2   &   2.4  $\pm$   0.6  &   19.8  $\pm$   5.9\\
Minni\,  8  &  1.35  &  9.50  &  14.7 $\pm$ 2.8 &   -0.9  &    0.9 $\pm$ 0.4   &   5.1 $\pm$ 1.3  &   42.8 $\pm$ 12.8    & disc/halo\\ 
          &        &        &7.2   &         &    0.4  $\pm$   0.2  &    2.5  $\pm$   0.6  &   20.9  $\pm$   6.3\\
Minni\,  9  &  1.60  &  8.40  &  10.3 $\pm$ 2.1 &   -0.1  &    0.6 $\pm$ 0.3   &   3.6 $\pm$ 0.9  &   30.0 $\pm$ 9.0     &bulge\\ 
          &        &        &8.5    &        &    0.5  $\pm$   0.2   &   3.0  $\pm$   0.7  &   24.7  $\pm$   7.4\\
Minni\, 10  &  1.70  &  9.40  &  15.6 $\pm$ 3.1  &  -0.3  &    0.9 $\pm$ 0.5   &   5.4 $\pm$ 1.4  &   45.4 $\pm$ 13.6    & ---\\ 
          &        &        & 9.5   &         &    0.6  $\pm$   0.3  &    3.3 $\pm$    0.8  &   27.6 $\pm$    8.3\\
Minni\, 11  &  1.80  &  8.70  &  11.7 $\pm$ 2.2  &  -0.3  &    0.7 $\pm$ 0.3   &   4.1 $\pm$ 1.0  &   34.0 $\pm$ 10.2    & ---\\ 
          &        &        &5.9    &        &    0.3  $\pm$   0.2  &    2.1 $\pm$    0.5  &   17.2   $\pm$  5.1\\
Minni\, 12  &  1.95  &  8.00  &   8.5 $\pm$ 1.7 &    +0.0  &    0.5 $\pm$ 0.2   &   3.0 $\pm$ 0.7  &   24.7 $\pm$ 7.4    &bulge\\ 
          &        &        &5.6    &        &    0.3 $\pm$    0.2   &   2.0  $\pm$   0.5  &   16.3  $\pm$   4.9\\
Minni\, 13  &  2.20  &  8.40  &   8.4 $\pm$ 1.6  &  -0.3  &    0.5 $\pm$ 0.2   &   2.9 $\pm$ 0.7  &   24.4 $\pm$ 7.3     &bulge\\ 
          &        &        &6.2    &        &    0.4 $\pm$    0.2  &    2.2 $\pm$    0.5  &   18.0  $\pm$   5.4\\
Minni\, 14  &  3.70  &  8.60  &  11.7 $\pm$ 2.5  &   0.0  &    0.7 $\pm$ 0.3   &   4.1 $\pm$ 1.0  &   34.0 $\pm$ 10.2    & ---\\ 
          &        &        &6.3    &         &   0.4  $\pm$   0.2   &   2.2 $\pm$    0.5  &   18.3  $\pm$   5.5\\
Minni\, 15  &  2.40  &  8.50   &  9.6 $\pm$ 1.9  &  -0.3  &    0.6 $\pm$ 0.3   &   3.4 $\pm$ 0.8  &   27.9 $\pm$ 8.4     &bulge\\ 
          &        &        &7.0    &         &   0.4  $\pm$   0.2   &   2.4  $\pm$   0.6  &   20.4  $\pm$   6.1\\
Minni\, 16  &  2.50 &   8.30   & 10.2 $\pm$ 2.1  &  -0.3  &    0.6 $\pm$ 0.3   &   3.6 $\pm$ 0.9   &  29.7 $\pm$ 8.9     &bulge\\ 
          &        &        &7.0    &        &    0.4  $\pm$   0.2   &   2.4  $\pm$   0.6  &   20.4  $\pm$   6.1\\
Minni\, 17  &  1.35  &  8.50  &   9.9 $\pm$ 1.9  &  -0.3  &    0.9 $\pm$ 0.3   &   3.5 $\pm$ 0.9  &   28.8 $\pm$ 8.6     &bulge\\
          &        &        &6.0    &         &   0.5  $\pm$   0.2   &   2.1  $\pm$   0.5  &   17.5  $\pm$   5.2\\
Minni\, 18  &  2.05  &  8.20  &  10.2 $\pm$ 2.1  &  +0.1  &    0.6 $\pm$ 0.3   &   3.6 $\pm$ 0.9  &   29.7 $\pm$ 8.9     &bulge\\
          &        &        &7.9    &         &   0.5  $\pm$   0.2   &   2.8  $\pm$   0.7  &   23.0  $\pm$   6.9\\
Minni\, 19  &  2.20  &  8.10  &  12.0 $\pm$ 2.6  &  +0.1  &    0.7 $\pm$ 0.3   &   4.2 $\pm$ 1.0  &   34.9 $\pm$ 10.5    & ---\\ 
          &        &        &8.1   &          &   0.5  $\pm$   0.2   &   2.8 $\pm$    0.7 &    23.6  $\pm$   7.1\\
Minni\, 20   & 2.45  &  6.90  &   7.6 $\pm$ 1.7  &  +0.2  &    0.4 $\pm$ 0.2   &   3.1 $\pm$ 0.7  &   22.1 $\pm$ 6.6     &bulge\\ 
          &        &        &7.3   &         &    0.4 $\pm$    0.2  &    3.0  $\pm$   0.6  &   21.2 $\pm$    6.4\\
Minni\, 21  &  1.80  &  8.60  &  12.4 $\pm$ 2.5  &  -0.3  &    1.1 $\pm$ 0.4   &   3.6 $\pm$ 1.1  &   32.5 $\pm$ 10.8    & ---\\ 
          &        &        &7.6  &          &    0.7 $\pm$    0.2   &   2.2  $\pm$   0.7  &   19.9  $\pm$   6.6\\
Minni\, 22  &  2.00  &  8.30  &  10.8 $\pm$ 2.2  &  +0.0  &    0.6 $\pm$ 0.3   &   3.8 $\pm$ 0.9  &   28.3 $\pm$ 9.4     &bulge\\
          &        &        & 6.6   &         &    0.4 $\pm$    0.2  &    2.3 $\pm$    0.6 &    17.3  $\pm$   5.8\\
\hline
\end{tabular}

\noindent Note: to convert 1 arcmin to pc, we use the following expression, $d$sin(1/60), where $d$ is
the derived heliocentric distance in pc.
\end{table*}

The slope of GC RGBs has been found to be related to the GC metallicity
\citep{valentietal04,cohenetal2015}. Indeed, at different fixed 
absolute magnitudes the mean colour of the RGB is different,
in the sense that the brighter the fixed RGB magnitude the larger the colour.
Although it is not highly sensible to metallicity
changes as compared with combinations involving blue and near-infrared passbands 
\citep[see, e.g.][]{getal97}, we still attempted some rough estimates under the assumption
that we are dealing with old GCs. To do this, we 
simply superimposed a subsample of theoretical isochrones computed by \citet{betal12} 
covering the metallicity range -1.5 dex $\le$ [Fe/H] $\le$ +0.6 dex on the $P$=100\%+$P$=88\% CMDs,
once they were corrected by distance and reddening effects.  
We then interpolated the
[Fe/H] values for the GC RGBs at $M_{K_s}$ = -5.5 mag, with an estimated
uncertainty of $\sigma$[Fe/H] $\approx$ 0.2 dex. Likewise, we confirmed 
that the
observed RGBTs are in excellent agreement with the positions suggested by the
theoretical isochrones. The resulting values are listed in 
Table~\ref{tab:table1}.

\subsection{Stellar density profiles}

 We built cluster radial density profiles using stars with different $P$ values in order 
to confirm that the RGB stars seen in the $P$= 100\% CMDs correspond to real stellar overdensities,
and that such overdensities progressively disappear as smaller $P$ values are used. 
We started by tracing annuli of 0.5  arcmin wide centred on the GC candidates, distributed
uniformly from their centres out to 10 arcmin. Each annulus was split in four equal area regions, 
which were used to count the number of stars placed inside them. We then averaged the 
four independent number of stars counted per annulus and produced an averaged observed stellar 
radial profile with its scatter being represented by the standard deviation. Fig.~\ref{fig:fig1} 
illustrates some resulting normalised stellar radial profiles for different $P$ values with their 
corresponding uncertainties.  As can be seen, these radial profiles show that
the stars with  the largest  $P$ values  belong to a stellar overdensity.

\begin{figure}
\includegraphics[width=\columnwidth]{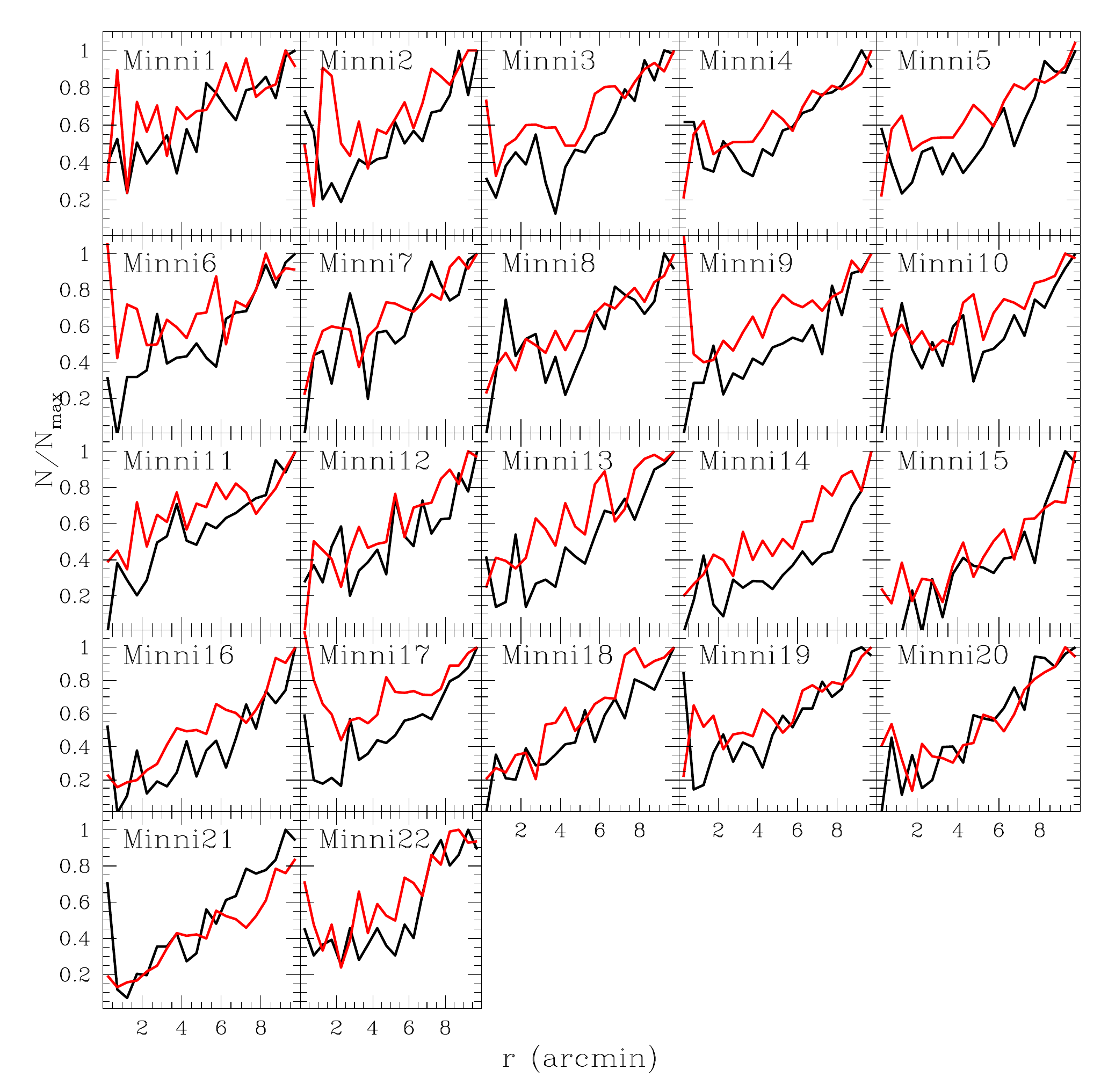}
    \caption{Normalised observed stellar radial profiles for stars with $P$ = 13\% and 25\% drawn with solid
black and red lines, respectively, for the 22 GC candidates.
}
   \label{fig:fig2}
\end{figure}

\begin{figure}
\includegraphics[width=\columnwidth]{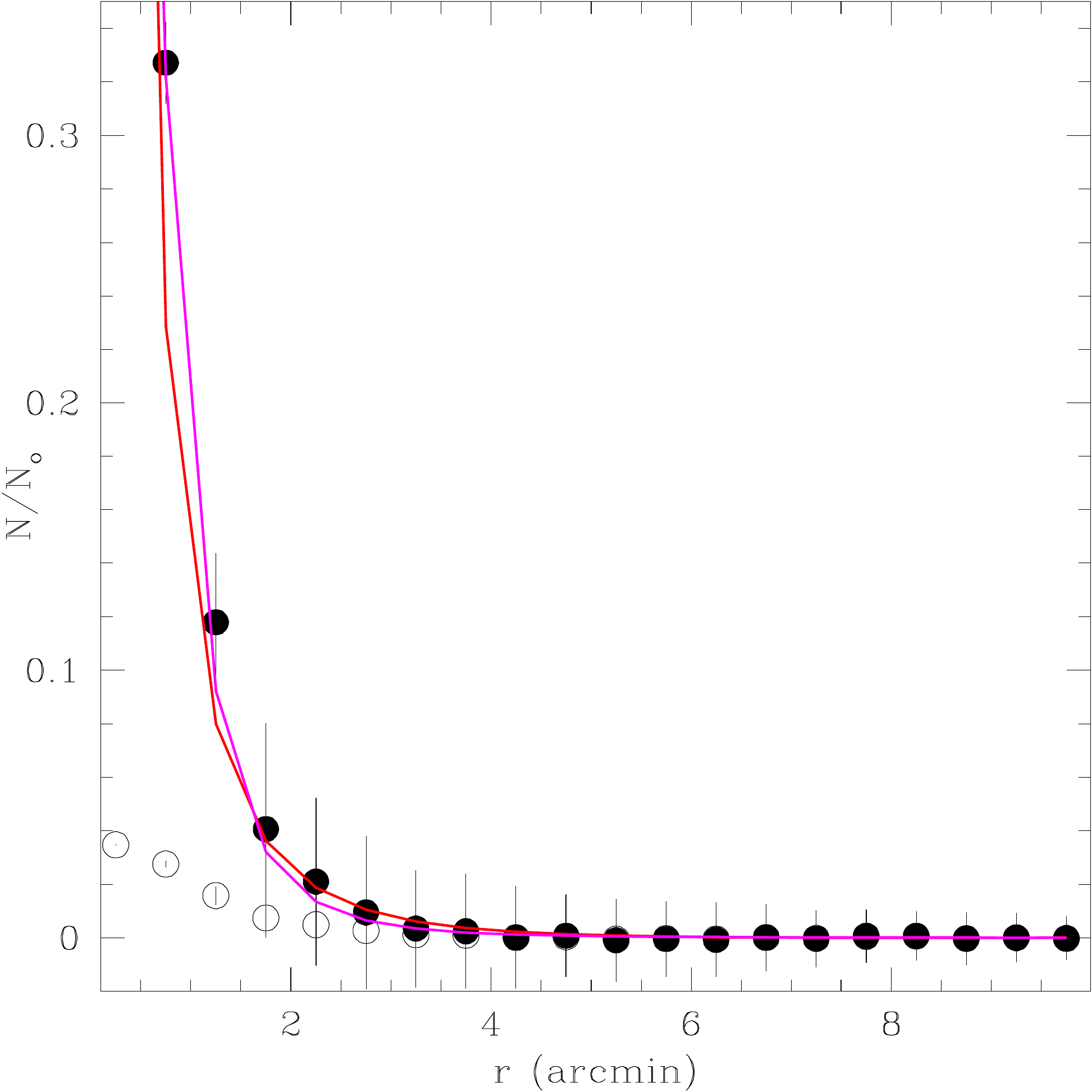}
    \caption{Normalised observed and corrected from crowding effect stellar radial profile of NGC\,6637
depiected with open and filled circles, respectively. Both distributions are background subtracted radial profiles.
Errorbars are also included. The red and magenta
solid lines represent the \citet{king62}'s and \citet{plummer11}'s models for the core, half-light and tidal radii compiled by \citet{harris1996} and \citet{vanderbekeetal2015}.
}
   \label{fig:fig3}
\end{figure}

Although 2MASS photometry is nearly 100\% complete down to $K_s$ $\sim$ 16 mag \citep{setal2006},
it suffers from incompleteness towards the inner GC clusters regions because of crowding effects.
In order to establish the actual radial dependence of the photometry incompleteness, we used
the information coming from our membership probability procedure. Stars with $P$ = 13\% and 25\%
should exhibit a nearly flat distribution of their observed radial profiles, because they
mainly represent the composite star field. However, 
Fig.~\ref{fig:fig2} shows that the closer to the GC candidates' centres, the smaller the number of counted
stars, which a is direct evidence of crowding effect. Therefore, we used all these radial profiles
to derive the mean relation $f$ = 0.099$r$ + 0.01, that we used to correct the $P$ = 88\%
and 100\% stellar radial profiles of the 22 GC candidates for crowding effects. 
The correction is carried out by dividing the observed radial profile by the function $f$.
In order to assess 
the reliability of such a correction, we built the stellar radial profile of NGC\,6637 as described 
above. We then corrected the resulting observed radial profile and overplotted on it the 
\citet{king62}'s and \citet{plummer11}'s models for the core ($r_c$), half-light ($r_h$) and tidal ($r_t$) radii
compiled by  \citet{harris1996} and \citet{vanderbekeetal2015}. Fig.~\ref{fig:fig3} confirms that
the derived correction satisfactorily restored the intrinsic NGC\,6637's radial profile.

We then corrected the GC candidate observed radial profiles from crowding effect, subtracted from
them the respective mean background level - measured from the outermost regions of the traced profiles -, 
and finally we normalised them. Subsequently, we fitted \citet{king62}'s models from a grid of
$r_c$ and $r_t$ values, as well as \citet{plummer11}'s profiles in order to have 
independent estimates
of the $r_h$ radii. The $r_c$, $r_h$, and $r_t$ values which made both models best 
resembled the GC candidate intrinsic profiles were derived from $\chi$$^2$  minimisation.
Fig.~\ref{fig:fig4} shows the normalised corrected radial profiles with the King and Plummer curves
superimposed, while Table~\ref{tab:table1} lists the resulting  $r_c$, $r_h$, and $r_t$ values
with their respective uncertainties. Below our values for each GC candidate, we included the 
M17's distance and the corresponding radii, for comparison purposes.

\begin{figure*}
\includegraphics[width=\columnwidth]{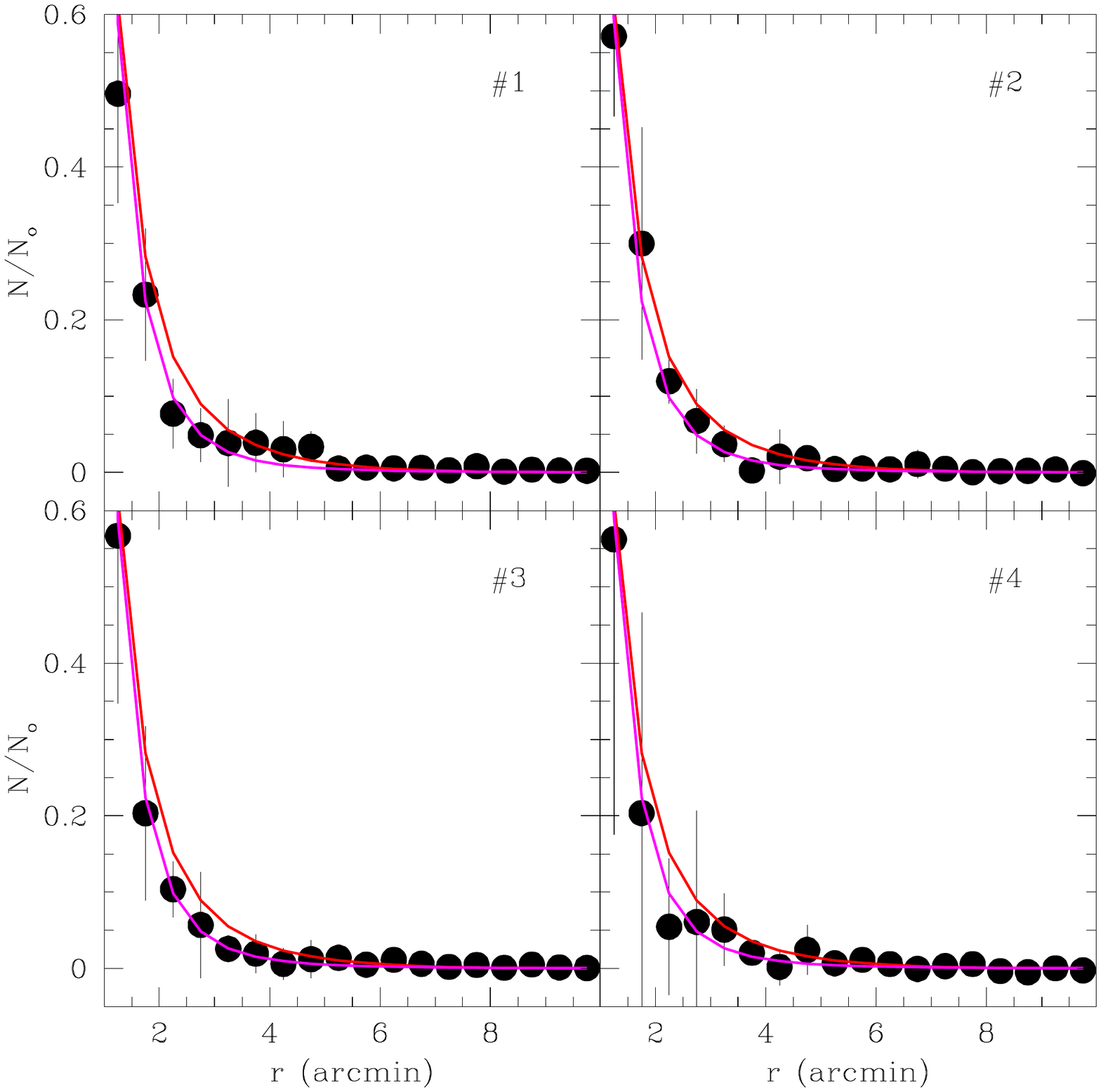}
\includegraphics[width=\columnwidth]{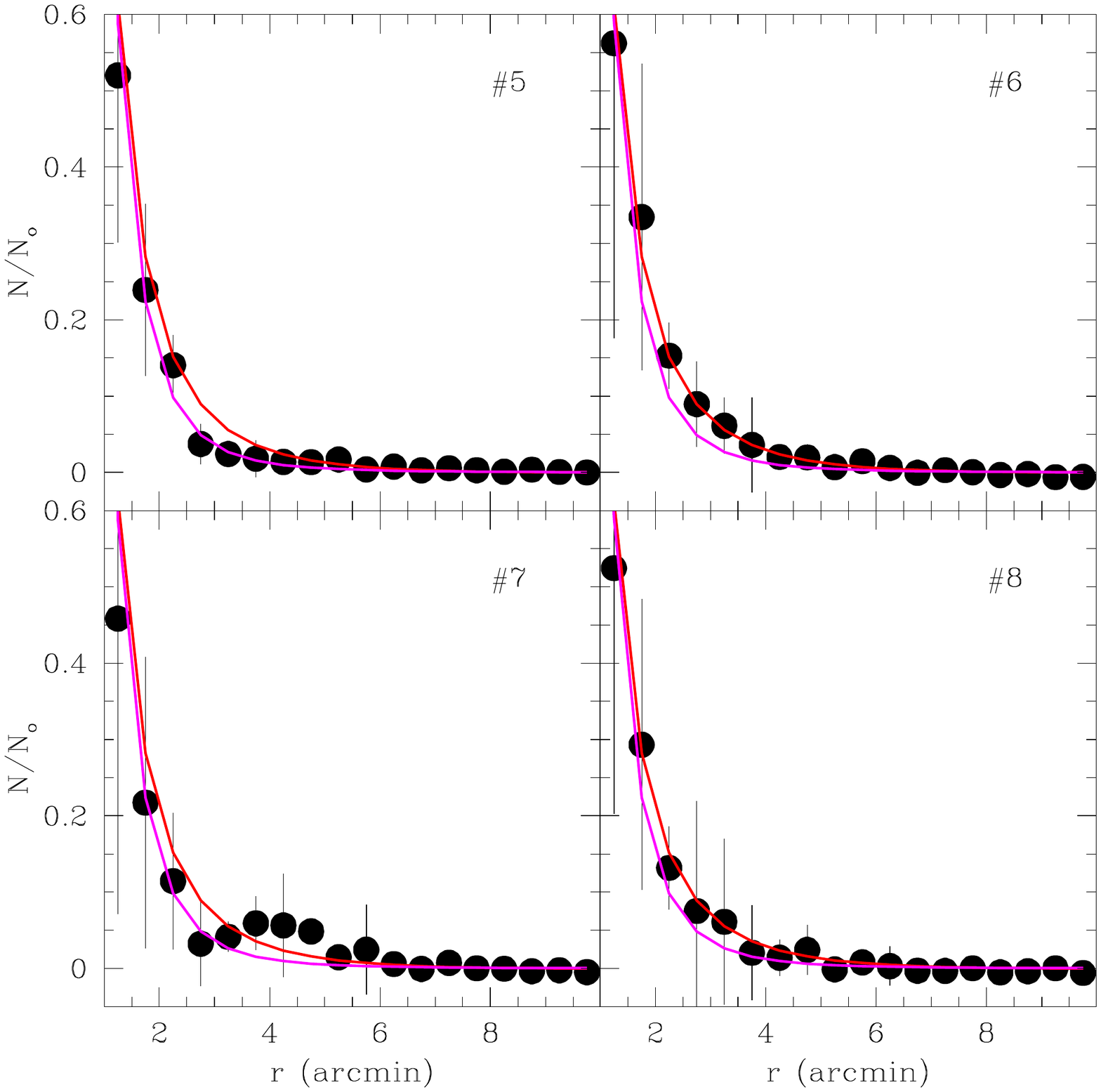}
\includegraphics[width=\columnwidth]{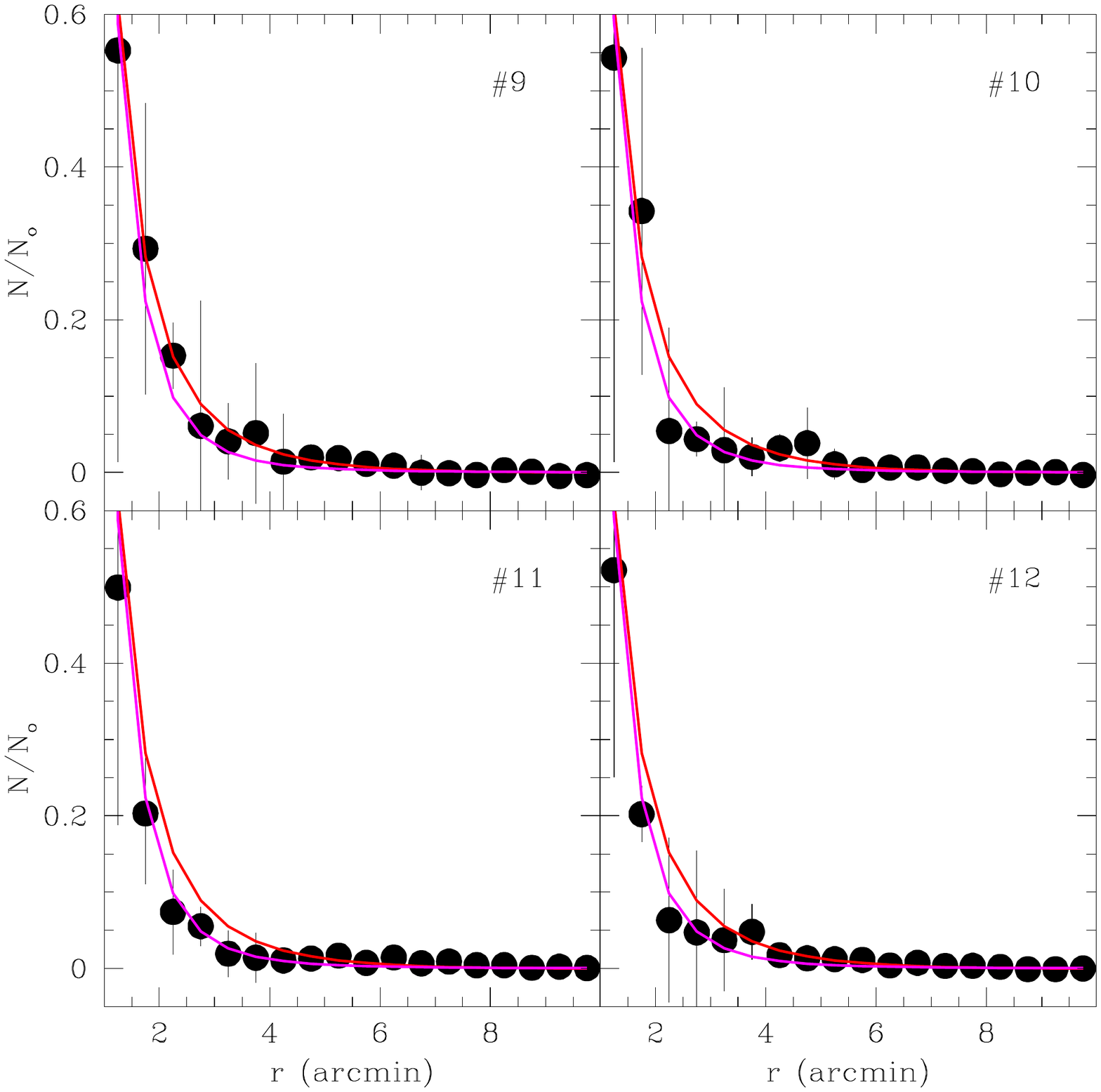}
\includegraphics[width=\columnwidth]{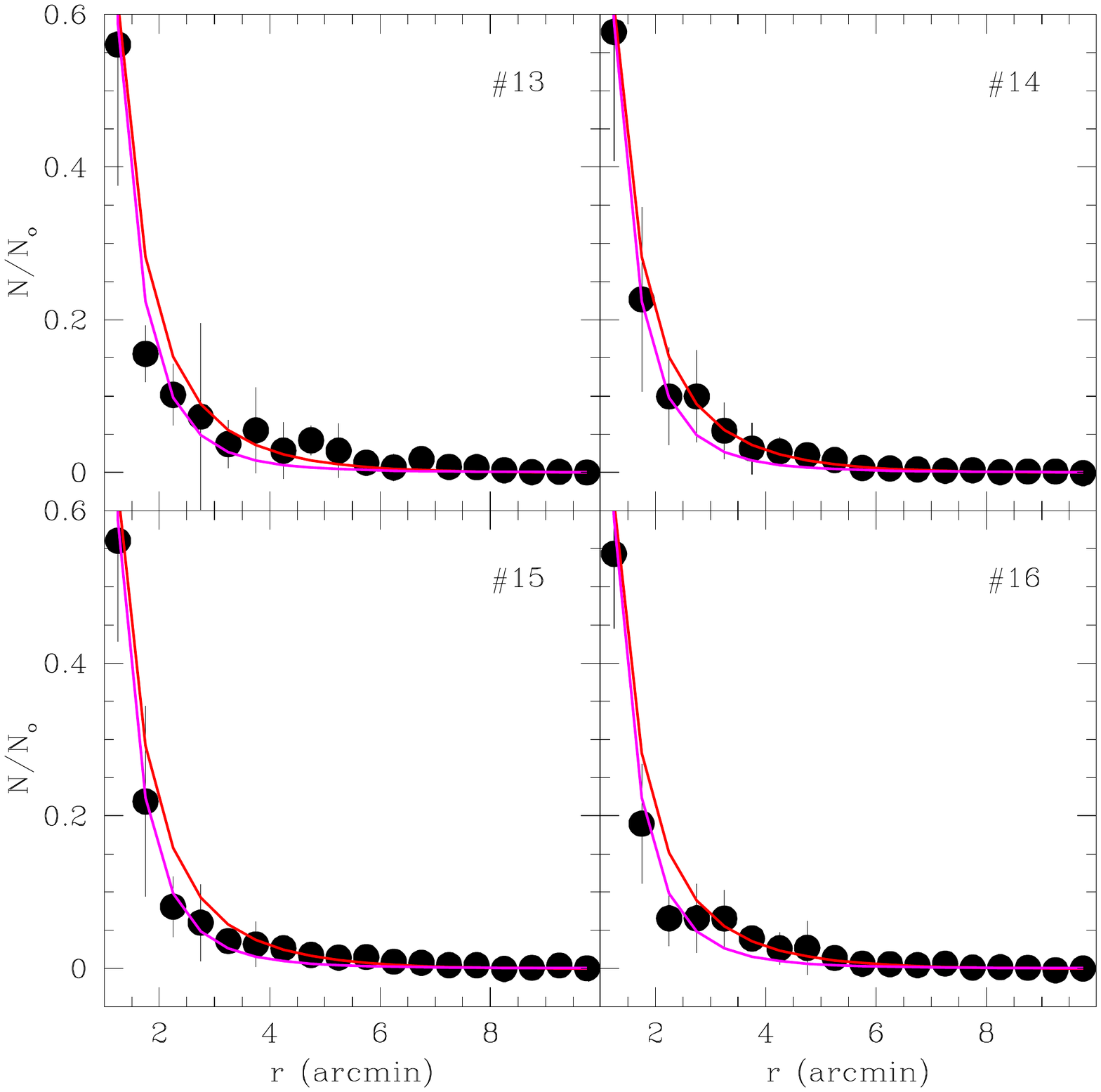}
    \caption{Normalised, corrected from crowding effects and background subtracted stellar radial
profiles for the studied GC candidates, with their respective errorbars. Solid red and magenta 
lines represent the fitted \citet{king62}'s and \citet{plummer11}'s models.
}
   \label{fig:fig4}
\end{figure*}

\setcounter{figure}{3}
\begin{figure*}
\includegraphics[width=\columnwidth]{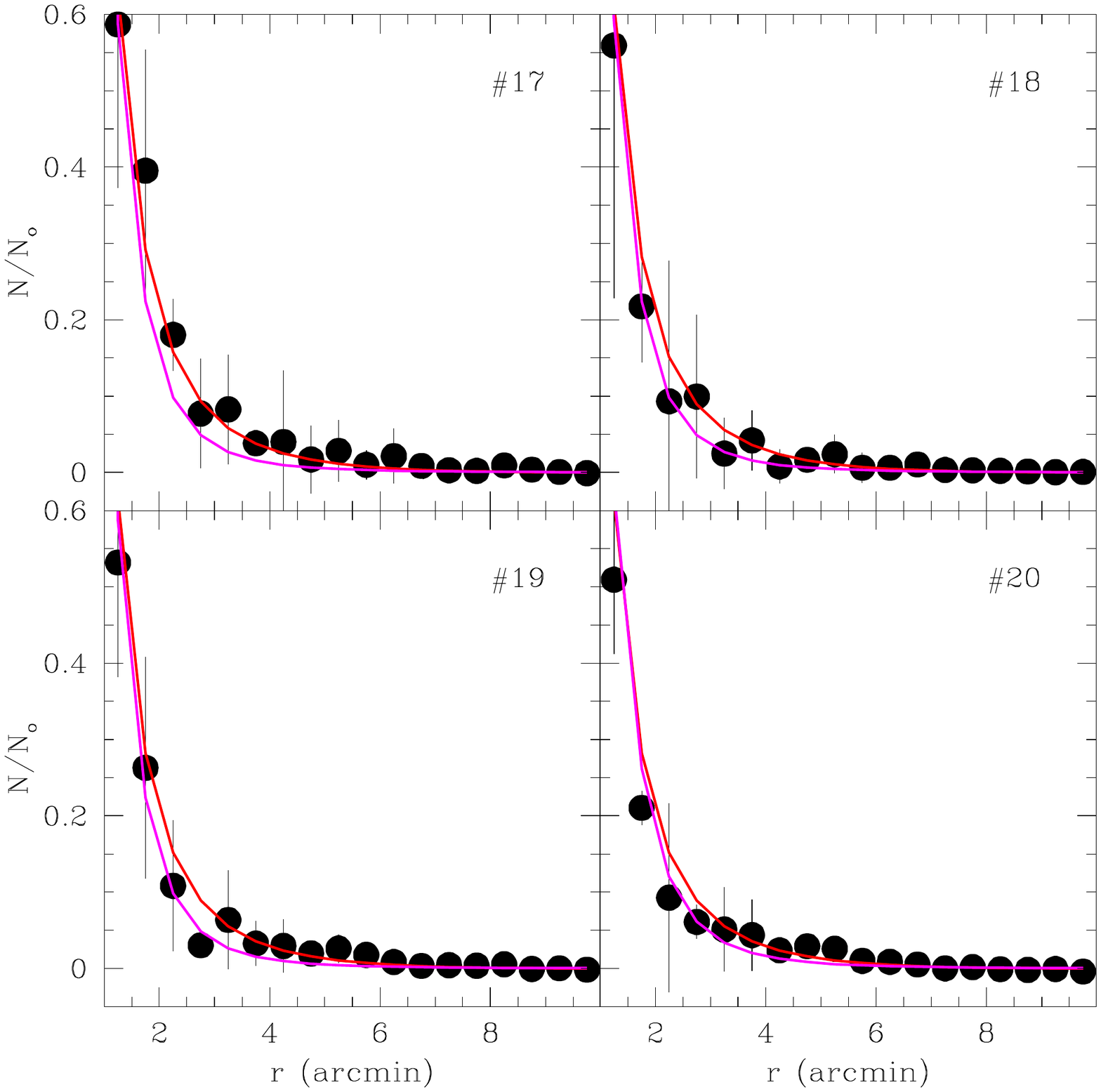}
\includegraphics[width=\columnwidth]{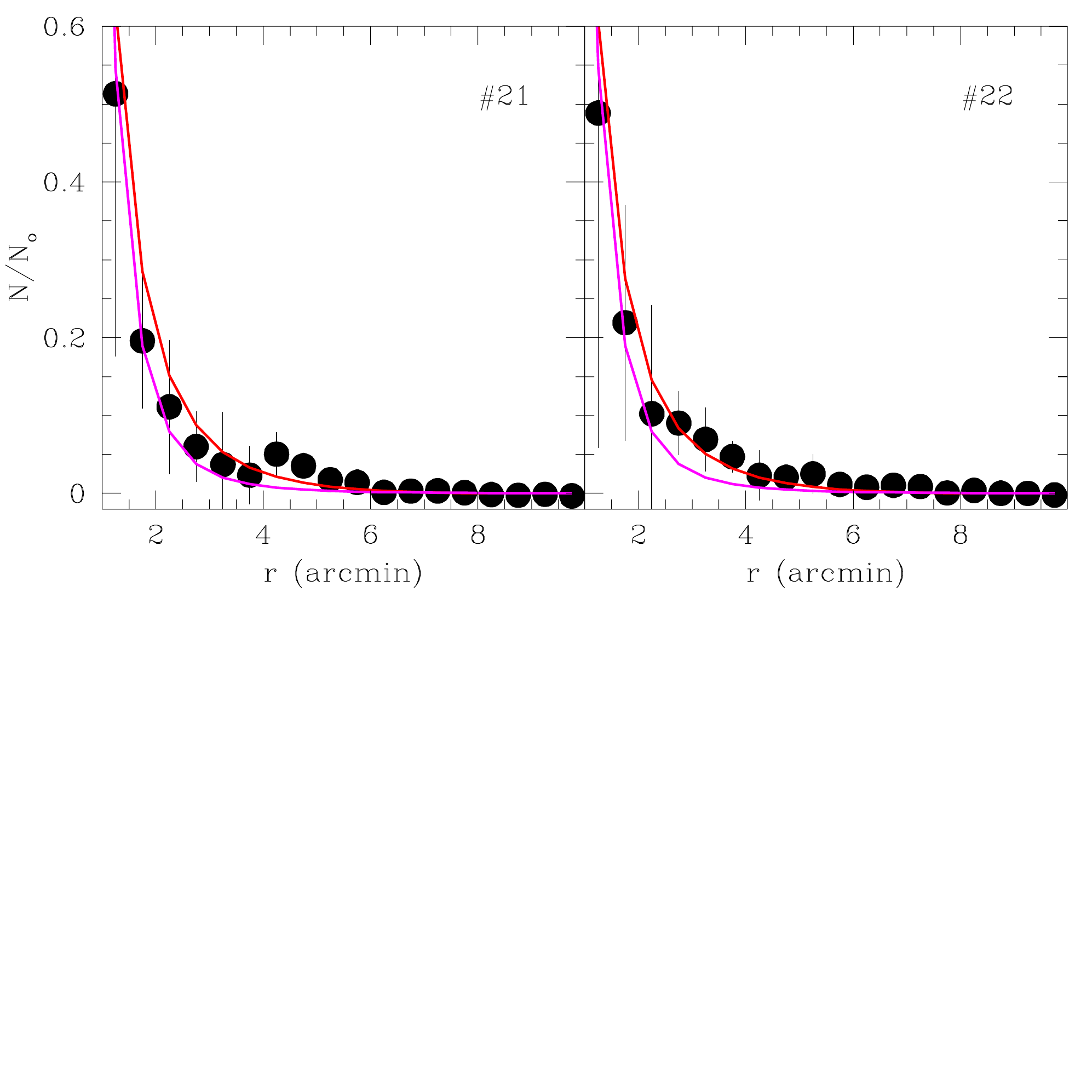}
    \caption{continued. 
}
   \label{fig:fig4}
\end{figure*}

\section{Analysis and discussion}

\begin{figure*}
\includegraphics[width=\textwidth]{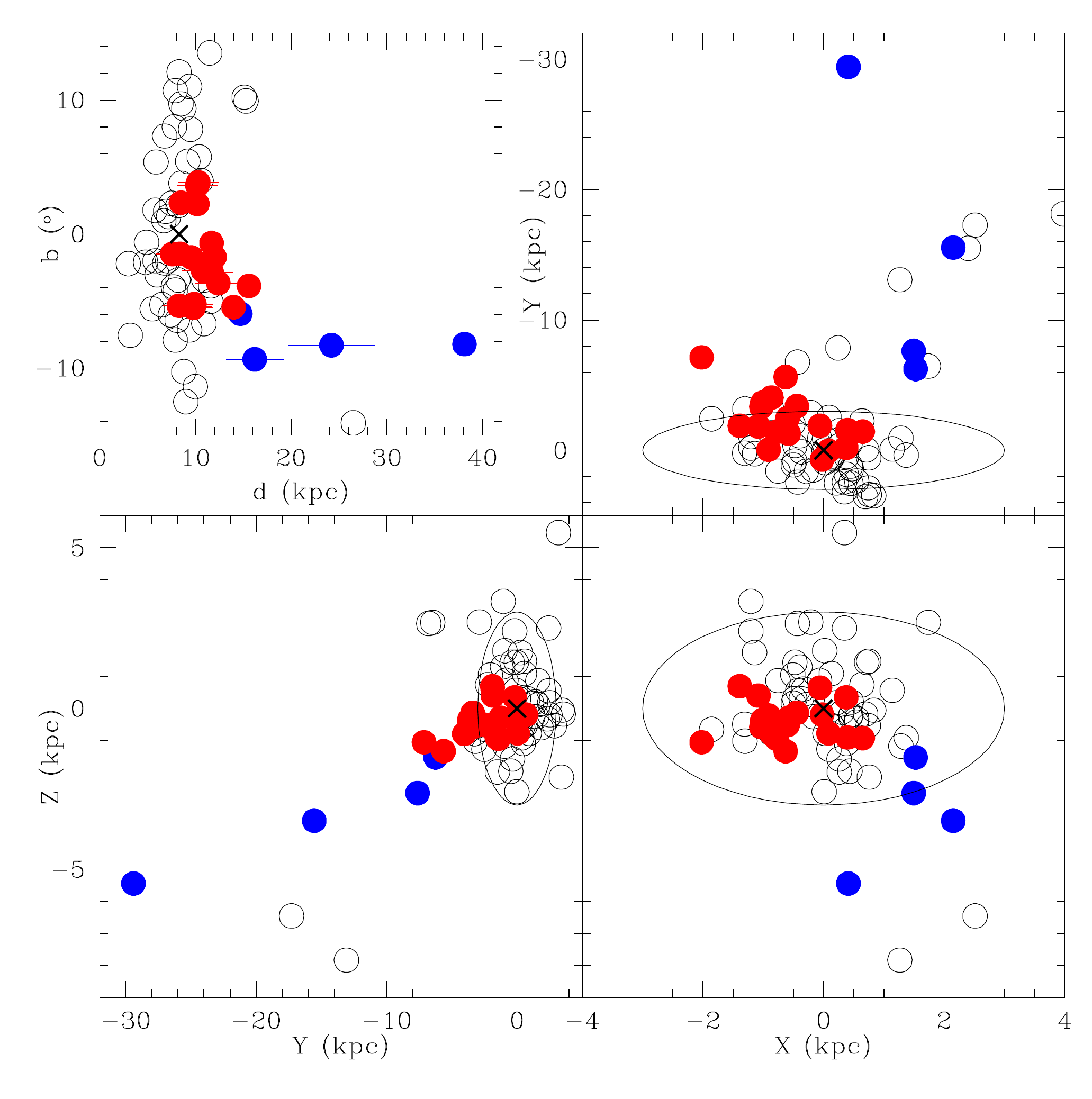}
    \caption{Spatial distribution of the studied GC candidates and those compiled by 
\citet{harris1996} for the same Galactic longitude interval drawn with filled and open
circles, respectively. Red and blue filled circles correspond to GC candidates with
metallicities smaller and larger than [Fe/H] = -0.8 dex. A cross represents the
position of the Galactic centre;  while the Sun is located  at 
 (b,d) = (0,0) and (X,Y,Z)$_{\odot}$=(0,8.3,0). A circle of radius 3 kpc is also
shown.}
   \label{fig:fig5}
\end{figure*}

We plotted in Fig.~\ref{fig:fig5} the distribution of the estimated
distances in different projected MW planes. Galactic coordinates were
calculated from the distances ($d$) listed in Table~\ref{tab:table1} 
and the Galactic coordinates ({\it l},b) provided in M17. The $X$, $Y$, $Z$
directions were chosen so that the $X$ axis increases from the Galactic centre
towards {\it l} = 270$\degr$, the $Y$ axis runs positive towards {\it l} = 
180$\degr$ and the $Z$ axis is perpendicular to the Galactic plane, pointing 
to the North Galactic pole. We used a Sun's distance to the Galactic centre
of 8.3 kpc \citep{dgb2017}. We distinguished GC candidates more metal-poor and
more metal-rich than [Fe/H] = -0.8 dex with blue and red filled circles,
respectively.  For  comparison purposes, we included the
positions of known MW GCs compiled by \citet{harris1996} with Galactic
longitudes in the same range as those for the GC candidates. They are
drawn with open circles.

The different projections reveal that many GC candidates are placed in the 
MW bulge, occupying regions where there were not any GC catalogued
previously. This is an interesting result, because these GC candidates
fill other parts of the MW bulge volume, giving thus a more complete picture of their
actual distribution. Note that \citet[][and references therein]{barrosetal2016}
describe the MW bulge as an oblate spheroid with a radius of 3 kpc encompassing
$\sim$ 90 per cent of its mass. Interestingly enough results also that
some GC candidates appear to be MW halo/disc objects, as judged by their
derived heliocentric distances \citep{minniti1995,cote1999,bicaetal2016}. 
Indeed, Minni\,3 seems to be one of the farthest
GCs to the Sun observed across the MW bulge so far. 

\begin{figure}
\includegraphics[width=\columnwidth]{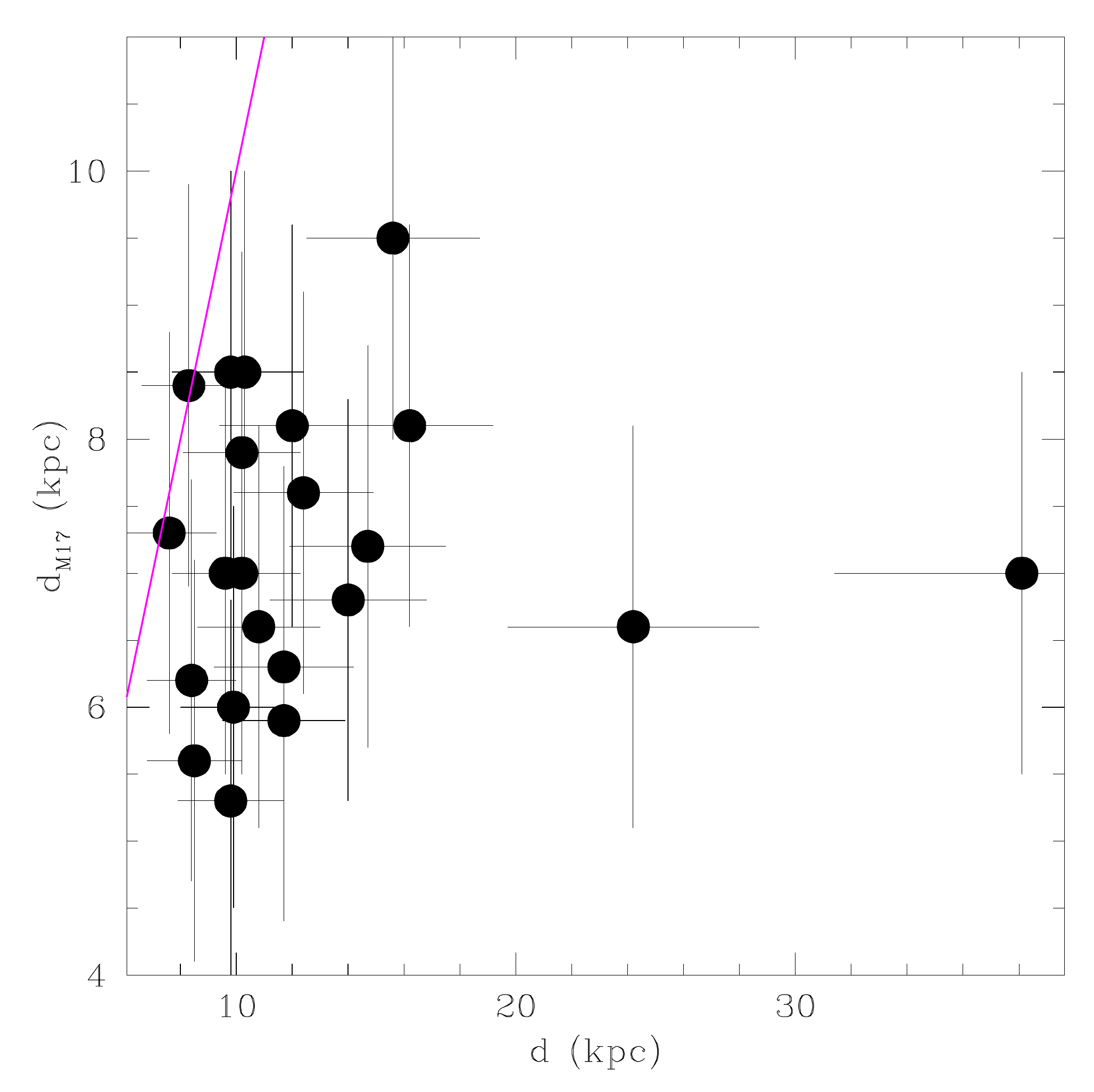}
    \caption{Comparison between heliocentric distances derived by M17 and in this work.
The magenta line represents the identity relationship.}
   \label{fig:fig6}
\end{figure}

The distinction between halo/disc and bulge GC candidates is supported by their estimated
metallicities. As can be seen, MW bulge GCs are more metal-rich
objects in comparison to those placed in the MW halo/disc 
\citep{minniti1995,cote1999,bicaetal2016}.
The present GC candidates spatial distribution is more consistent with their
physical nature as MW bulge GCs - except those located out of the bulge -  than that
that comes from considering the heliocentric distances estimated by M17. Indeed,
a simple comparison shows that their heliocentric distances are in average $\sim$ 4 kpc
smaller than ours (see Fig.~\ref{fig:fig6}). 

Besides Galactic positions and metallicities, structural parameters 
such as core, half-light and tidal radii can also
help us to disentangle the physical reality of the studied objects. 
They can tell us about the GC candidates internal dynamical histories. 
For instance, some MW GCs are core-collapsed objects or clusters in a highly 
advanced dynamical stage. This implies that their concentration parameters 
$c$ ($\equiv$ log($r_t$/$r_c$)) are  within the largest possible values.
 This is because two-body relaxation mechanisms
cause outermost cluster regions expand while the cluster core squeezes. Galactic 
tidal forces
also play an additional role, frequently producing  an extra expansion of the
outermost cluster regions \citep[see, e.g.][and references therein]{p17c,p18}.

\begin{figure*}
\includegraphics[width=\textwidth]{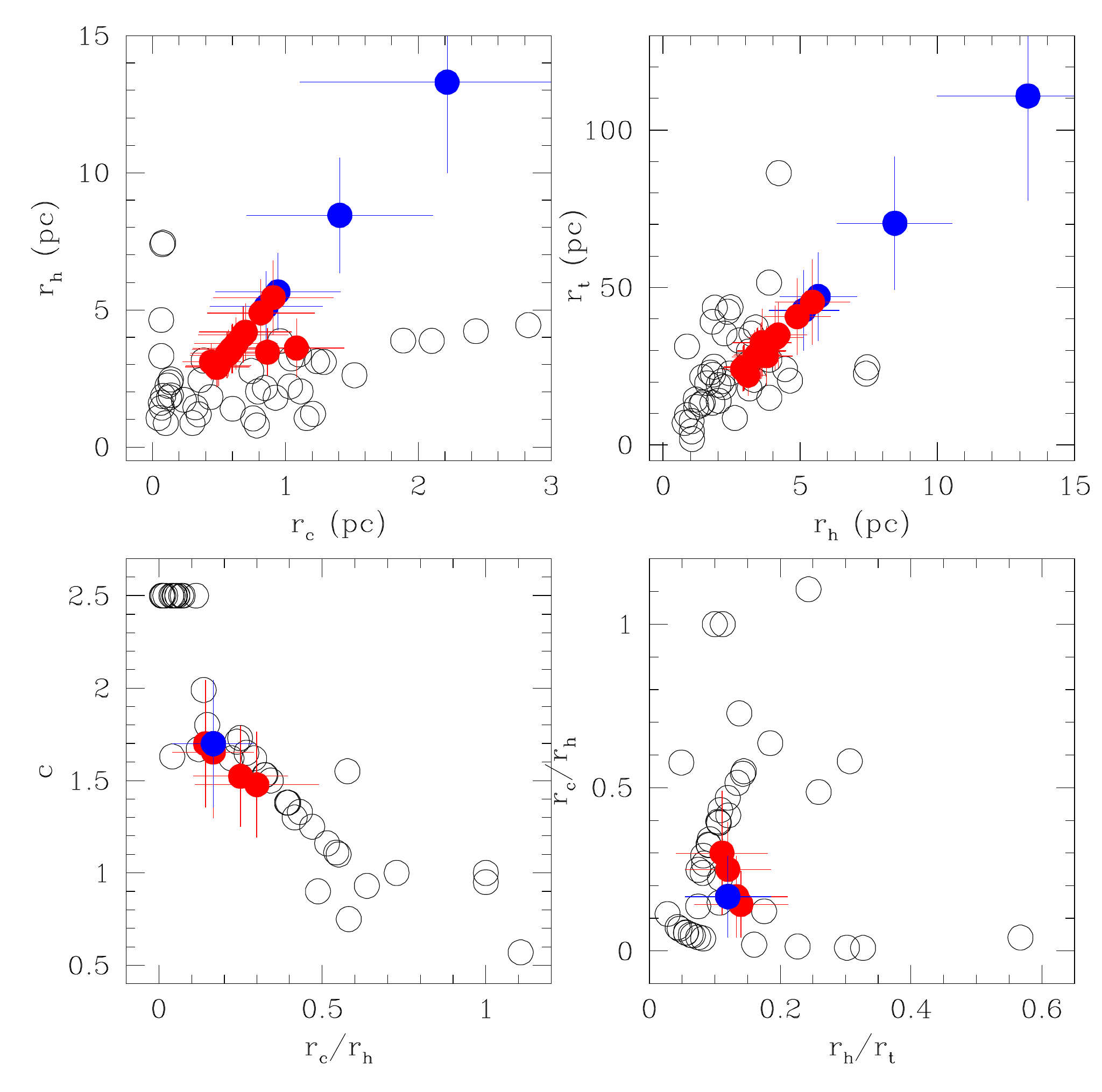}
    \caption{Relationships between different structural parameters with their
respective errorbars. Symbols are as in Fig.~\ref{fig:fig5}.
}
   \label{fig:fig7}
\end{figure*}

We started by comparing our estimated structural parameters ($r_c$, $r_h$ 
and $r_t$) with those of MW bulge GCs. We restricted the comparison to this 
particular subsample in order to facilitate the analysis with the
bulge GC candidates. We assumed that if the studied bulge GC candidates are
genuine bulge GCs, they should have structural parameters in the ranges
spanned by already known MW bulge GCs. Fig.~\ref{fig:fig7} depicts different
structural parameter relationships for them. The upper panels show that the
$r_c$, $r_h$, $r_t$ values of the studied bulge GC candidates are comparable with
those of catalogued MW bulge GCs. In the figure we have represented
studied and catalogued GCs with filled and open circles. The bulge GC 
candidates are drawn with red circles, while halo/disc GC candidates 
- included for completeness purposes - are shown with blue circles. 
 As can be seen, there 
is a general superposition between the different structural parameter ranges,
although the studied GC candidates tend to cover the regime with larger 
$r_h$ values.
It would be interesting to trace stellar radial profiles for these objects
from a much higher spatial resolution photometry, in order to probe whether
the incompleteness correction applied from stellar crowding effects could
lead to slightly larger values, particularly in the case of $r_h$. As
for $r_t$, we think that they should not be residually affected by this correction.

The bottom panels of Fig.~\ref{fig:fig7} shows that the ratio between
different structural parameters are in fairly good agreement with those for 
known MW bulge GCs.  According to \citet{harris1996}, concentration parameters 
close to $c$ $\sim$ 2.5 are core-collapsed GCs. They also appear at the bottom-left
corner of the $r_c/r_h$ vs $r_h/r_t$ plane,   where a cluster 
moves approximately in the top-right-bottom-left direction  
\citep[][see, e.g., their figure 33.2]{hh03} while relaxing toward a core-collapse stage.
The stage of internal dynamical evolution of the studied GC candidates depend on their
masses and on the MW tidal field as well. Although we could assume that MW bulge 
GCs have spent most of their lifetime describing small orbits around the MW centre, and
therefore they have been similarly affected by the gravitational field, we would need 
to know their masses in order to compute half-light relaxation times. With this latter
property we would be able to describe more properly their real dynamical stages.
Table~\ref{tab:table1} lists in the last column the classification of the studied
GC candidates on the basis of their heliocentric distances and metallicities as 
discussed above. We did not assign any classification to six 
GCs with [Fe/H] $\ge$ -0.8 dex located beyond a circle of
3 kpc from the Galactic centre, because their metallicities and heliocentric distances
lead to different classifications.

\section{Conclusions}

From 2MASS $JK_s$ photometry we have addressed the issue about the physical
entity of recently reported 22 GC candidates located towards the MW bulge,
from the  VVV survey. The analysis performed is  twofold. 
On the one hand, we used different data sets,
and on the other hand, the CMD features examined and the methodology employed
to estimate astrophysical properties are
different as well. Here we took advantage of the RGBT as a distance estimator.

We relied our analysis on an extensively applied and successful procedure to
assign photometric membership probabilities to stars measured
in the GC candidate field. Thus, we were able to disentangle the composite
star field population from those real local stellar enhancements around the
GC candidates central positions. Such stellar overdensity  does not only
constitute a genuine distinct group of stars projected on the observed field,
but also their RGBs are remarkably well-delineated and narrower than those for
the composite star field population. Both results led us to confirm their
nature as stellar aggregates. 

Their estimated heliocentric distances place
 them behind the Galactic centre; 
several of them near to bulge regions where no previous GC had been identified, 
whereas some few other objects
resulted to be MW halo/disc clusters. The distinction between MW bulge and halo/disc
GC candidates is fully supported by our estimated metallicities.

We derived for the first time their structural parameters, namely: core, half-light and
tidal radii. The different resulting radii suggest that these GC candidates do not
differentiate from the population of known MW GCs, and particularly from those of
the bulge. They are comprised within the typical bulge GC dimensions and within
the expected dynamical evolutionary stages.

\section*{Acknowledgements}

This publication makes use of data products from the Two Micron All Sky Survey, 
which is a joint project of the University of Massachusetts and the Infrared 
Processing and Analysis Center/California Institute of Technology, funded by 
the National Aeronautics and Space Administration and the National Science Foundation.
 We thank the referee for his/her thorough reading of the manuscript and
timely suggestions to improve it. 

%%%%%%%%%%%%%%%%%%%%%%%%%%%%%%%%%%%%%%%%%%%%%%%%%%



\bibliographystyle{mnras}
%\bibliography{paper} % if your bibtex file is called paper.bib

%to be uncommented before sending to editor
\input{paper.bbl}

%\begin{thebibliography}{99}
%\end{thebibliography}

%%%%%%%%%%%%%%%%%%%%%%%%%%%%%%%%%%%%%%%%%%%%%%%%%%

%%%%%%%%%%%%%%%%% APPENDICES %%%%%%%%%%%%%%%%%%%%%

%\appendix


%%%%%%%%%%%%%%%%%%%%%%%%%%%%%%%%%%%%%%%%%%%%%%%%%%

% Don't change these lines
\bsp	% typesetting comment
\label{lastpage}
\end{document}